\documentclass[9pt]{article}

\usepackage{geometry}
\usepackage{microtype}
\usepackage{subfigure}
\usepackage{booktabs} 

\usepackage{multirow}
\usepackage{graphicx}
\usepackage{amsmath,amsthm}
\theoremstyle{definition}
\newtheorem{definition}{Definition}
\newtheorem{problem}{Problem}
\usepackage{xspace,color}
\usepackage{diagbox,multirow} 
\usepackage{subfigure}
\usepackage{tikz,bm}
\usepackage{algorithmic,algorithm}
\definecolor{PineGreen}{rgb}{0.0, 0.47, 0.44}
\usepackage{enumitem}
\setlist[itemize]{leftmargin=*}
\usepackage{amsfonts}
\usepackage[utf8]{inputenc}
\usepackage[T1]{fontenc}    
\usepackage{url}            
\usepackage{booktabs}       
\usepackage{nicefrac}       
\usepackage{microtype}      
\usepackage{tikz,bm}
\usepackage{adjustbox}      
\usepackage{filecontents}
\usepackage{array,rotating}
\usepackage[labelfont=bf]{caption}

\usepackage{caption}
\usepackage{xspace,color}
\usepackage{enumerate}
\usepackage{enumitem,amsmath,amssymb}
\usepackage{hyperref}

\newcommand{\RB}{\mathbb{R}}

\newcommand{\bfe}{\mathbf{e}}

\newcommand{\bfh}{\mathbf{h}}

\newcommand{\bfp}{\mathbf{p}}

\newcommand{\bfu}{\mathbf{u}}

\newcommand{\bfs}{\mathbf{s}}
\newcommand{\bfS}{\mathbf{S}}
\newcommand{\bfz}{\mathbf{z}}

\newcommand{\Wf}{\mathbf{W}}

\newcommand{\haty}{\hat{y}}
\newcommand{\calA}{\mathcal{A}}
\newcommand{\calD}{\mathcal{D}}
\newcommand{\calE}{\mathcal{E}}
\newcommand{\calG}{\mathcal{G}}

\newcommand{\calL}{\mathcal{L}}
\newcommand{\calM}{\mathcal{M}}

\newcommand{\calT}{\mathcal{T}}

\newcommand{\mbf}[1]{{\boldsymbol{\mathbf{#1}}}}
\renewcommand{\bm}{\mbf}

\newcommand{\marked}{}

\newcommand{\mname}{\texttt{HINT}\xspace}
\newcommand{\data}{\texttt{TOP}\xspace}
\newcommand{\stdfontsize}{\footnotesize}
\newcommand{\fullname}{Hierarchical Interaction Network\ }

\usepackage{amssymb,bbm}
\usepackage{amsmath,amsthm,amsfonts,booktabs}
\usepackage[switch]{lineno}

\theoremstyle{definition}

\begin{document}
%
\title{\mname: \fullname for Clinical Trial Outcome Prediction}


\author{Tianfan Fu$^{1*}$ \and Kexin Huang$^{2*}$ \and Cao Xiao$^3$ \and Lucas~M.~Glass$^{3,4}$ and Jimeng Sun$^5$}
\date{%
    $^1$Georgia Institute of Technology\\
    $^2$Harvard University \\
    $^3$Analytics Center of Excellence, IQVIA\\
    $^4$Temple University\\
    $^5$University of Illinois at Urbana-Champaign\\ [2ex]%
    Feb 02, 2021\footnote{The latest version of the paper is accepted by Cell Patterns, available at \url{https://www.sciencedirect.com/science/article/pii/S2666389922000186}.  }
}
\maketitle

\begin{abstract}
Clinical trials are crucial for drug development but  
are time consuming, expensive, and often burdensome on patients.  More importantly, clinical trials face uncertain outcomes due to issues with efficacy, safety,  or problems with patient recruitment. If we were better at predicting the results of clinical trials, we could avoid having to run trials that will inevitably fail — more resources could be devoted to trials that are likely to succeed.
In this paper, we propose Hierarchical INteraction Network  (\mname) for more general, clinical trial outcome predictions for all diseases based on a comprehensive and diverse  set of web data including molecule information of the drugs, target disease information,  trial protocol and biomedical knowledge. 
 \mname first encode these multi-modal data into latent embeddings, where  an imputation module is designed to handle missing data. Next,  these embeddings will be fed into the  knowledge embedding module to generate knowledge embeddings that are pretrained using external knowledge on pharmaco-kinetic properties and  trial risk from the web. Then the interaction graph module will connect all the embeddings via domain knowledge to fully capture various trial components and their complex relations as well as their influences on trial outcomes. Finally, \mname  learns a dynamic attentive graph neural network to predict trial outcome. 
Comprehensive experimental results show that \mname achieves strong predictive performance, obtaining 0.772, 0.607, 0.623, 0.703 on PR-AUC for Phase I, II, III, and indication outcome prediction, respectively. It also consistently outperforms the best baseline method by up to 12.4\% on PR-AUC. 
The curated dataset and code repository is available at \url{https://github.com/futianfan/clinical-trial-outcome-prediction}. 
The latest version of the paper is accepted by Cell Patterns, available at \url{https://www.sciencedirect.com/science/article/pii/S2666389922000186}. 
\end{abstract}

\section{Introduction}
Clinical trial is an indispensable step towards developing a new drug, where human participants are tested in responding to a treatment (e.g., a drug or drug combinations) for treating target diseases. The costs of conducting clinical trials are extremely expensive (up to hundreds of millions of dollars~\cite{martin2017much}) and the time to run a trial is very long with low success probability~\cite{peto1978clinical,ledford20114}.
However, many factors such as the inefficacy of the drug, drug safety issues, and poor trial protocol design can cause the failure of a clinical trial~\cite{friedman2015fundamentals}. If we were better at predicting the results of clinical trials, we can avoid running trials that will inevitably fail — more resources could be devoted to trials that are more likely to succeed. Fortunately, the online availability of historical clinical trial data, and massive knowledge  bases about approved and failed drugs bring a new opportunity for using machine learning models to tackle the key question: {\it Can one predict the success probability of a trial before the trial started?}

Various data sources on the web can provide important knowledge for predicting the trial outcome. For example, clinicaltrials.gov database\footnote{Publicly available at \url{https://clinicaltrials.gov/}.} has 279K clinical trial records, which describe many important information about clinical trials. In addition, we are able to find the standard medical codes for most of the diseases described in natural language through National Institutes of Health website\footnote{Publicly available at \url{https://clinicaltables.nlm.nih.gov/}. }. DrugBank Database\footnote{Publicly available at \url{https://www.drugbank.ca/}}~\cite{wishart2018drugbank} contains the biochemical description of many drugs, which allows the computational modeling of drug molecules. 


Over the years, there have been early attempts in predicting individual components in clinical trials to improve the trial results, including using electroencephalographic (EEG) measures to predict the effect of antidepressant treatment in improving depressive symptoms ~\cite{raj20evaluation}, optimizing drug toxicity based on drug-property and target-property features~\cite{hong20predicting}, leveraging phase 2 results to predict the phase 3 trial results~\cite{pmlr-v106-qi19a}. Recently, there has been interest in developing a general method for trial outcome prediction. As an initial attempt, \cite{Lo2019Machine} expanded  beyond optimizing individual component to predict trial outcomes for 15 disease groups based on disease-only features using statistical modeling.
Despite these initial efforts, there are several limitations that impede the utility of existing trial outcome prediction models. 


\begin{itemize}[leftmargin=*]
\item \textbf{Limited task definition and study scope.} Existing works either focus on predicting individual component of  trials~\cite{hong20predicting,gayvert2016data,artemov2016integrated,dong2018admetlab} such as patient-trial matching or only covering disease groups of which the disease-specific features are available~\cite{Lo2019Machine}. Although these works are potentially useful for a limited part of the trial design, they do not answer the fundamental problem: will this trial succeed? To the best of our knowledge, there is no work that attempts to solve the general trial outcome prediction problem across different trial phases for many different diseases.

\item \textbf{Limited features used for prediction.} Due to their limited task definition and study scope, existing works often only leverage restricted disease-specific features, which cannot be  generalized for other diseases. These works also ignore the facts that trial outcomes are determined by various trial risks including drug safety, treatment efficiency and trial recruitment, where abundant information exist on the web. For example, the biomedical knowledge that provide explicit biochemical structures among drug molecules and previous trials history for certain disease are both ignored by existing studies but deem very useful for trial outcome prediction. 
\item \textbf{Failed to explicitly capture the complex relations among trial components and trial outcomes.} Due to the limited data and task scope, existing methods often simplify 
their predictions by limited input features and use simple computational method that is not explicitly designed for trial outcome prediction~\cite{hong20predicting,gayvert2016data,artemov2016integrated,pmlr-v106-qi19a,Tranchevent750364}. This significantly impedes them to model the complicated relations among various trial components.

\end{itemize}


\noindent\textbf{Our Approach}.
To provide accurate trial outcome prediction for all trials, we propose a \fullname (\mname).  \mname has an  input embedding module  to encode  data from various web sources including drug molecules, disease information and trial protocols, where  an imputation module is designed to handle missing data. Next,  these embeddings are fed into the  knowledge embedding module to generate knowledge embeddings that are pretrained using external knowledge on pharmaco-kinetic properties and  trial risk from the web. Then the  interaction graph module will connect all the embeddings via domain knowledge to fully capture various trial components and their complex relations as well as their influences on trial outcomes. Based on that, \mname  learn a dynamic attentive graph neural network to predict trial outcome. 

\smallskip
\noindent\textbf{Contribution}. Our main contributions are listed as follows:
\begin{itemize}[leftmargin=*]
\item {\bf Problem} We formally define a model framework for a general clinical trial outcome prediction task, which not only models various trial risks including drug safety, treatment efficiency and trial recruitment,  but also models a wide range  of drugs and indications (e.g., diseases). Our model framework can generalize over new trials given the drug, disease and protocol information (Section~\ref{sec:formulation}). 
\item {\bf Benchmark} To enable general clinical trial outcome predictions, we leverage a comprehensive set of datasets from various public web sources, including drug bank, standard disease code and clinical trial records and curate a public-available clinical trial outcome prediction dataset \data. This benchmark dataset will be released and open-sourced to the community to advance machine learning-aided clinical trial design soon after the review process (Section~\ref{sec:data}).
\item {\bf Method} We design a machine learning method that explicitly simulates each clinical trial component and the complicated relations among them (Section~\ref{sec:overview}-\ref{sec:imputation}). 
\end{itemize}

We evaluated \mname against state-of-the-art baselines using real world data. 
\mname achieved 0.772, 0.607, 0.623 with PR-AUC on Phase I, II, III level prediction respectively and 0.703 on indication-level prediction. These high absolute scores suggest the practical usage of \mname in predicting clinical trial outcome in various stages of the trials. In addition, \mname has up to 12.4\% relatively improvement in terms of PR-AUC compared with best baseline method (COMPOSE)~\cite{gao2020compose}. We also conduct ablation study to evaluate the importance of key components of clinical trials to the prediction power and the effectiveness of the hierarchical formulation of a trial interaction graph. At last, we conduct a case study to show the potential real world impact of \mname by successfully predicting the failure of some recent huge trial efforts.

\section{Related Work}
\label{sec:related}

\subsection{Trial Outcome Prediction}

Existing works often focus on predicting individual patient's outcome in a trial instead of a general prediction about the overall trial success. They usually leverage expert crafted features. For example, \cite{wu2012identifying} leveraged SVM to predict the status of genetic lesions based on cancer clinical trial documents. 
\cite{raj20evaluation} use Gradient-Boosted Decision Trees (GBDT~\cite{ye2009stochastic}) to predict the improvement in symptom scores based on treatment symptom score and electroencephalographic (EEG) measures for depressive symptoms with antidepressant treatment. 
\cite{hong20predicting} focus on predicting clinical drug toxicity according to drug-property and target-property features and use an ensemble classifier of weighted least squares support vector regression. Note that these models are not tackling the same task as us. They are predicting in a patient-level whereas \mname focuses on trial-level. More relevant to us, \cite{pmlr-v106-qi19a} designs Residual Semi-Recurrent Neural Network (RS-RNN) to predict the phase 3 trial results based on phase 2 results. In contrast, the task of \mname is to predict for all clinical trial phases. \cite{Lo2019Machine} explored various imputation techniques and a series of conventional machine learning models (including logistic regression, random forest, SVM) to predict the outcome of clinical trial within 15 disease groups. However, they do not consider drug molecule features and trial protocol information and thus could not differentiate the outcome for different drugs focusing on a disease, and cannot capture trial failure due to poor enrollment whereas \mname takes account into all of these information. 
 
\subsection{Trial Representation Learning}
Recently, deep learning has been leveraged to learn representation from clinical trial data to support downstream tasks such as patient retrieval~\cite{zhang2020deepenroll,gao2020compose} and enrollment~\cite{Biswal2020Doctor2VecDD}. For example, Doctor2Vec~\cite{Biswal2020Doctor2VecDD} learn hierarchical clinical trial embedding where the unstructured trial descriptions were embedded using BERT \cite{devlin2018bert}. DeepEnroll~\cite{zhang2020deepenroll} also leveraged a Bidirectional Encoder Representations from Transformers (BERT~\cite{devlin2018bert}) model to encode clinical trial information. More recently, COMPOSE~\cite{gao2020compose} used pretrained BERT to generate contextualized word embedding for each word of trial protocol, and then applied  multiple one-dimensional convolutional layers with different kernel sizes to generate trial embedding in order to capture semantics at different granularity level. While these works optimize the representation learning for a single component in a trial, \mname is the first to model a diverse set of trial components such as molecule, disease, protocols, pharmaco-kinetics, and  disease risk information and fuse all of them through a unique hierarchical interaction graph.


\section{Method}
\label{sec:method}

\subsection{Problem Formulation}
\label{sec:formulation}
A \underline{\textit{clinical trial}} is designed to validate  the safety and efficacy of a  \underline{\it treatment set}  towards a \underline{\it target disease set} on a patient group defined by the   \underline{\it trial protocol}.

\begin{definition}[\textbf{Treatment Set}]
Treatment set includes one or multiple drug candidates, denoted by
\begin{equation}
\label{eqn:molecule}
\mathbb{M} = \{{m}_1, \cdots, {m}_{N_m}\},  
\end{equation} 
where $m_1, \cdots, m_{N_m}$ are $N_m$ drug molecules involved in this trial.
\end{definition} 
Note that we focus on clinical trials that aim at discovering new indications of drug candidates. Other trials that do not involve molecules such as surgeries and devices are out of scope and can be considered as future work.

\begin{definition}[\textbf{Target Disease Set}]
Each trial targets at one or more diseases. 
Suppose there are $N_d\geq 1$ diseases in a trial, we represent the target disease set as 
\begin{equation}
\label{eqn:disease}
\mathbb{D} = \{d_1,\cdots, d_{N_d}\}, 
\end{equation}
where $d_1,\cdots, d_{N_d}$ are $N_d$ diseases. We use $d_i$ to represent the raw information associated with the disease including the disease name, description (text data) and its corresponding diagnosis code (e.g., International Classification of Diseases - ICD codes~\cite{anker2016welcome}). 
\end{definition}

Each trial has a trial protocol (in unstructured natural language) that describes eligibility criteria for enrolling patients including participants characteristics such as age, gender, medical history, target disease conditions, and current health status.

\begin{definition}[\textbf{Trial Protocol}]
The patient group is specified by the trial protocol. 
Formally, each protocol consists of a set of inclusion and exclusion criteria for recruiting patients, which describe what are desired and unwanted from the targeted patients, respectively, 
\begin{equation}
\label{eqn:criteria}
\quad \mathbb{C} = [\bm{c}_{1}^{I}, ..., \bm{c}_{M}^{I}, \bm{c}_{1}^{E}, ..., \bm{c}_{N}^{E}],\ \ \ \ \ \ \bm{c}^{I/E}_{i} \ \text{is a sentence}. 
\end{equation}
$M$ ($N$) is the number of inclusion (exclusion) criteria in the trial, $\bm{c}_{i}^{I}$ ($\bm{c}_{i}^{E}$) denotes the $i$-th inclusion  (exclusion) criterion. 
Each criterion $\bm{c}$ is a sentence in unstructured natural language. 
\end{definition}



\begin{problem}[\textbf{Trial Outcome Prediction}]
The trial outcome is a binary label $y\in\{0,1\}$, where $y=1$ indicates trial success and 0 indicates trial failure. 
The predicted success probability is  $\haty \in [0,1]$. The goal of \mname is to learn a deep neural network model $f_{\theta}$ for predicting the trial outcome $y$:
\begin{equation}
y = f_{\theta}(\mathbb{M}, \mathbb{D}, \mathbb{C}), 
\end{equation}
where $\mathbb{M}, \mathbb{D}, \mathbb{C}$ are the treatment set, target disease set and trial protocol, respectively. We  predict trial outcomes under 2 settings:
\begin{itemize}[leftmargin=*]
\item \textbf{Phase Level Prediction} focuses on predicting the outcome for a particular phase of the trial. In general, there are three trial phases: phase I tests the toxicity and side effects of the drug; phase II determines the efficacy of the drug (i.e., if the drug works); phase III focus on the effectiveness of the drug (i.e., whether the drug is better than the current standard practice). 
The phase level prediction determines whether a specific clinical trial study will successfully complete at the phase. 
\item \textbf{Indication Level Prediction} aims at predicting whether the indication of a drug will be approved (i.e., the treatment for a disease will pass all 3 phases).
\end{itemize}
\end{problem}
For ease of exposition, we list mathematical notations in Table~\ref{table:notation}.

\begin{table}[tb]
\small 
\centering
\caption{Mathematical notations.  }
\resizebox{\columnwidth}{!}{
\begin{tabular}{l|c|l}
\toprule[1pt]
Categories & Notations & Explanations \\ 
\midrule
\multirow{4}*{\shortstack[l]{Web Data\\ \& Output \\ (Section~\ref{sec:formulation}) }} 
& $\mathbb{M} = \{{m}_1, \cdots, {m}_{N_m}\}$  & $N_m$ Drug molecules, Eq.~\eqref{eqn:molecule}.  \\
& $\mathbb{D} = \{d_1,\cdots, d_{N_d}\}$  &  $N_d$ Diseases, Eq.~\eqref{eqn:disease}. \\
& $\mathbb{C} = [\bm{c}_{1}^{I}, \cdots, \bm{c}_{M}^{I}, \bm{c}_{1}^{E}, \cdots, \bm{c}_{N}^{E}]$  & Protocol criteria, Eq.~\eqref{eqn:criteria}. \\
& $\haty \in [0,1]$ & Predicted trial success probability\\
\midrule
\multirow{4}*{\shortstack[l]{Input\\Embedding \\ (Section~\ref{sec:input})}}
 & $\bfh_m \in \RB^d$, $f_m()$ & drug embedding, drug encoder \\
 & $\bfh_d \in \RB^d$, $\text{GRAM}()$ & disease embedding, disease encoder\\
 & $\alpha_{ji}, g_1(\cdot), \bfe_{i/j/k}$ & GRAM parameters~\cite{choi2017gram}, Eq.~\eqref{eqn:gram}, \eqref{eqn:gram_attention}  \\
 & $\bfh_p \in\RB^d $, $f_p()$ & trial embedding, trial encoder\\
 \midrule
\multirow{7}*{\shortstack[l]{Embedding and\\ its function \\ (Section~\ref{sec:knowledge},\ref{sec:gnn})}}
& $\bfh_A, \bfh_D, \bfh_M, \bfh_E, \bfh_T \in\RB^{d}$ & embeddings for A,D,M,E,T, Eq.~\eqref{eqn:admet} .  \\ 
& and $\mathcal{X}_A, \mathcal{X}_D, \mathcal{X}_M, \mathcal{X}_E, \mathcal{X}_T$ & and corresponding embedding functions \\
& $\bfh_R \in\RB^{d}$ and $\mathcal{R}()$ &  Disease risk,Eq.~\eqref{eqn:risk}.   \\ 
& $\bfh_{PK}\in\RB^{d}$ and $\mathcal{K}()$  &  Pharmaco-Kinetics, Eq.~\eqref{eqn:pk}.  \\
& $\bfh_I \in\RB^{d}$ and $\mathcal{I}()$ & Interaction, Eq.~\eqref{eqn:interaction}.  \\
& $\bfh_V \in\RB^{d}$ and $\mathcal{V}()$ & Augmented interaction, Eq.~\eqref{eqn:ai} \\ 
& $\bfh_{\text{Pred}} \in\RB^{d}$ and $\mathcal{P}()$ &  Prediction, Eq.~\eqref{eqn:trial}.  \\ 
\midrule
\multirow{6}*{\shortstack[l]{Hierarchical \\ Interaction\\ Graph \\ (Section~\ref{sec:gnn})}} 
  & $\calG, |\calG| = K$ & Interaction graph with $K$ nodes. \\ 
  & $\mathbf{A} \in \{0,1\}^{K\times K}$ & Adjacency matrix of $\calG$, Eq.~\eqref{eqn:gcn} \\
  & $\mathbf{V}\in \RB_{+}^{K\times K}, g_2(\cdot)$ & Attentive matrix/function, Eq.~\eqref{eqn:gcn},~\eqref{eqn:attention}.   \\ 
  & $\mathbf{B}\in\RB^{K\times d}$ & Bias parameter of GNN. \\ 
  & $\mathbf{W}^{(l)}\in\RB^{d\times d},\ \ l=0,\cdots,L$ & Weights in the $l$-th layer \\   
  & $\mathbf{H}^{(l)}\in\RB^{K \times d}, \ \ l=0,\cdots,L$ &  the $l$-th layer node embedding \\ 
\midrule
\multirow{2}*{\shortstack[l]{Imputation \\ (Section~\ref{sec:imputation})}} 
  & $\widehat{\bfh_m} \in\RB^d$ & Recovered drug embedding \\ 
  & $\calL_{\text{Recovery}}$ &  Recovery loss, Eq.~\eqref{eqn:recovery_loss}.  \\
\bottomrule[1pt]
\end{tabular}}
\label{table:notation}
\end{table}

\subsection{Overview of \mname}
\label{sec:overview} 

\begin{figure*}[h!]
\centering
\includegraphics[width =\textwidth]{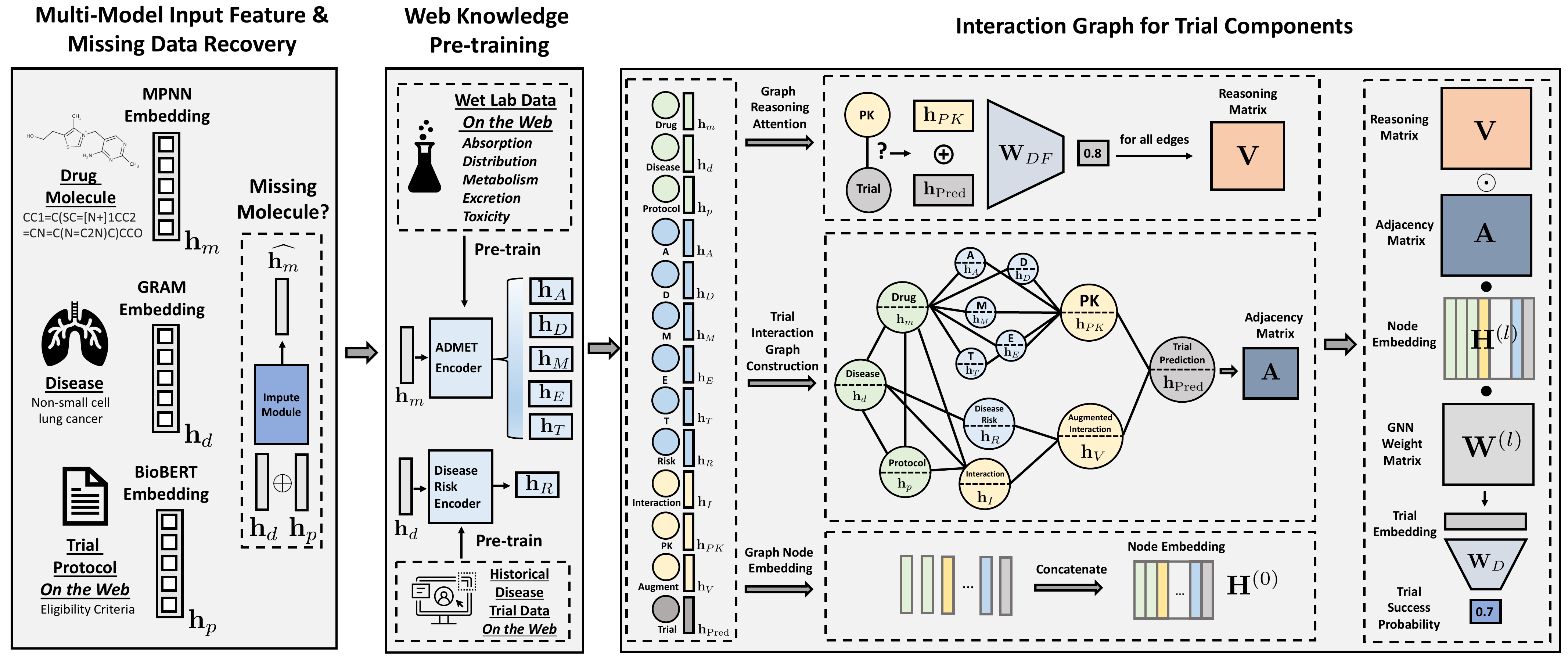}
\caption{\mname Framework. \mname takes features that describe the following trial components: drug molecule embeddings $\bfh_m$, disease embedding $\bfh_d$, and trial protocol embedding $\bfh_p$ (Section~\ref{sec:input}). Before constructing an interaction graph using these components, \mname pretrain some embeddings (blue nodes) using web based knowledge about drug properties and disease risks (Section~\ref{sec:knowledge}). Next, in Section~\ref{sec:gnn}, we construct an interaction graph to characterize interactions between various trial components. Trial embeddings are learned based on the interaction graph to capture both trial components and their interactions. Based on the learned representation and \textit{Dynamic Attentive Graph Neural Network} (Eq.~\ref{eqn:gcn}), we make trial outcome prediction. }
\label{fig:method}
\end{figure*}

As illustrated in Figure~\ref{fig:method},
\mname is an end-to-end framework for predicting the success probability of a trial before the trial starts. 
First, \mname has an {\bf input embedding module} to encode multi-modal data from various sources including drug molecules, disease information and trial protocols to input embeddings (Section~\ref{sec:input}). Next, these embeddings will be fed into the {\bf knowledge embedding module} to generate knowledge embeddings that are pretrained using external knowledge on pharmaco-kinetic properties and  trial risk from Web (Section~\ref{sec:knowledge}). Then the {\bf interaction graph module} will connect all the embeddings via domain knowledge to fully capture various trial components and their complex relations as well as their influences on trial outcomes. Based on that, \mname learn a \textit{dynamic attentive graph neural network} to predict trial outcome (Section~\ref{sec:gnn}). An {\bf imputation module} is designed to handle missing data (Section~\ref{sec:imputation}).

\subsection{Input Embedding Module}
\label{sec:input}

In this section, we describe the raw data and input representation learning.
Raw data used for trial outcome prediction come from three different sources, namely (1) drug molecules, (2) disease information and (3) trial protocols. 

\smallskip
\noindent{\bf 1. Drug molecules} are important for predicting trial outcomes. The drug molecules are represented by SMILES strings or molecule graphs. There are many existing works in embedding drug molecules into latent vectors including knowledge base approach such as Morgan fingerprint and its variants~\cite{cereto2015molecular}, or representation learning methods for SMILES strings~\cite{wang2019smiles} and molecule graphs~\cite{coley2017convolutional,fu2021differentiable,du2022interpretable,zhou2022network}. 
\mname supports all three types of molecule embeddings. Formally, we represent all molecules $\mathbb{M} = \{m_1, \cdots, m_{N_m}\}$ as drug (molecule) embedding
\begin{equation}
\label{eqn:molecule_feature}
\text{Drug (Molecule) Embedding}\quad \bfh_m = \frac{1}{N_m}\sum_{j=1}^{N_m} f_m(m_j),\ \ \bfh_m \in \RB^{d}. 
\end{equation}
where $f_m(\cdot)$ is the molecule embedding function and we average all molecule embeddings from a trial to get the {\it drug embedding vector}. 
Our experiments show that average is a better aggregation function than summation for drug embedding. 
The molecule embedding function $f_m$ can be any of the above drug embedding methods. \mname supports Morgan fingerprint~\cite{cereto2015molecular}, SMILES encoder~\cite{wang2019smiles}, graph message passing neural network (MPNN)~\cite{huang2020deeppurpose,fu2021mimosa} for creating the drug embedding. 

\smallskip
\noindent{\bf 2. Disease information} can also affect the trial outcome. 
For example, drugs in oncology have much lower approval rate than ones in infectious diseases~\cite{Hay2014-qw}. 
The disease information comes from its description and its corresponding ontology such as disease hierarchies like International Classification of Diseases (ICD)~\cite{anker2016welcome}.

Formally, we represent all diseases $\mathbb{D} = \{d_1,\cdots, d_{N_d}\}$ (Eq.~\ref{eqn:disease}) as disease embedding, 
\begin{equation}
\label{eqn:disease_feature}
\text{Disease Embedding}\quad\ \bfh_d = \frac{1}{N_d}\sum_{i=1}^{N_d} \text{GRAM}(d_i),\ \ \bfh_d \in \RB^{d}.  
\end{equation}
where $\text{GRAM}(d_i)$ represent an embedding of disease $d_i$ using GRAM (graph-based attention model)~\cite{choi2017gram}. 
GRAM is designed to leverage the hierarchical information inherent to medical ontologies and succesfully used in predictive phenotyping~\cite{choi2017gram,fu2021probabilistic}. Specifically, the representation of current disease $d_i$ is a convex combination of the basic embeddings ($\bfe\in \RB^{d}$) of itself and its ancestors, i.e, 
\begin{equation}
\label{eqn:gram}
\text{GRAM}(d_i) = \sum_{j\in \text{Ancestors}(i)\cup \{i\}} \alpha_{ji} \bfe_j, 
\end{equation}
where $\alpha_{ji}\in \RB_{+}$ represents the attention weight and we have 
\begin{equation}
\label{eqn:gram_attention}
\begin{aligned}
& \alpha_{ji} = \frac{\exp\big(g_1([\bfe_j^\top,\bfe_i^\top]^\top)\big)}{\sum_{k\in \text{Ancestors}(i)\cup \{i\}} \exp\big(g_1([\bfe_{k}^\top,\bfe_i^\top]^\top)\big)},\\ 
& \sum_{j\in \text{Ancestors}(i)\cup \{i\}} \alpha_{ji} = 1,
\end{aligned}
\end{equation}
where $g_1(\cdot)$ is a feed-forward network with a single hidden layer, following~\cite{choi2017gram}.  $\text{Ancestors}(i)$ represents the set of all the ancestors of $i$. 
An ancestor of code represent a higher-level category of the current code. 
For example, in ICD 10 code~\cite{anker2016welcome}, ``D41'' (urinary organs neoplasm) and ``D41.2'' (ureter neoplasm) are the ancestors of  ``D41.20'' (right ureter neoplasm). 
``C34'' (malignant neoplasm of bronchus and lung) and ``C34.9'' (malignant neoplasm of unspecified part of bronchus or lung) are the ancestors of ``C34.91'' (malignant neoplasm of unspecified part of right bronchus or lung).

\smallskip
\noindent{\bf 3. Trial protocol} is a document that describes how a clinical trial will be conducted, includes eligibility criteria, which describes the patient recruitment requirements. 
Eligibility criteria contain both inclusion and exclusion criteria denoted by Eq.~\eqref{eqn:criteria}, 
\[
\mathbb{C} = [\bm{c}_{1}^{I}, ..., \bm{c}_{M}^{I}, \bm{c}_{1}^{E}, ..., \bm{c}_{N}^{E}], \ \ \ \ \ \ \bm{c}^{I/E}_{i} \ \text{is a sentence}
\]
where $\bm{c}^I_i$ and $\bm{c}^E_j$ denote the $i$-th and $j$-th sentence in inclusion and exclusion criteria, respectively. To convert the criteria sentences into embedding vectors, we apply 
Clinical-BERT~\cite{alsentzer2019publicly,huang2019clinicalbert} which is a domain-specific version of BERT~\cite{devlin2018bert}. 
\begin{equation}
\label{eqn:clinicalbert}
\begin{aligned}
& \bfs^I_i = \text{Clinical-BERT}(\bm{c}^I_i),\ \  \bfS^I = [\bfs^I_1, \cdots, \bfs^I_M],\\
& \bfs^E_j = \text{Clinical-BERT}(\bm{c}^E_i),\ \  \bfS^E = [\bfs^E_1, \cdots, \bfs^E_N]. 
\end{aligned}
\end{equation}
Then we use 4 one-dimensional convolutional layers~\cite{you2018end,gao2020compose} (denoted Conv1d) with different kernel sizes ($k_1,\cdots,k_4$) to capture semantics at 4 different granularity level. Specifically, $\bfS^I$ and $\bfS^E$ are fed into four one-dimensional convolutional layers and the output are concatenated as 
\begin{equation}
\label{eqn:convolutional1}
\begin{aligned}
& \bfp^I = \text{CONCAT}\Big(\text{Conv1d}(\bfS^I, k_1),\cdots, \text{Conv1d}(\bfS^I, k_4)\Big) \\
& \bfp^E = \text{CONCAT}\Big(\text{Conv1d}(\bfS^E, k_1),\cdots, \text{Conv1d}(\bfS^E, k_4)\Big).
\end{aligned}
\end{equation}
Then a one-layer fully connected neural network (denoted by $\text{FC}(\cdot)$) is used to build the protocol embedding $\bfh_p$ based on the concatenation of $\bfp^I$ and $\bfp^E$,  
\begin{equation}
\label{eqn:protocol_embedding_last}
\bfh_p = \text{FC}(\text{CONCAT}(\bfp^I, \bfp^E)). 
\end{equation}
Thus, the protocol embedding is written as 
\begin{equation}
\label{eqn:protocol_embedding}
\text{Protocol Embedding}\quad\ \bfh_p = f_p(\mathbb{C}), \ \ \ \ \bfh_p \in \RB^{d}. 
\end{equation}

\subsection{Pretraining Using Web Knowledge}
\label{sec:knowledge}
\mname utilizes vast amount of web data to further enhance those input embeddings. 
In particular, \mname leverages drug bank data~\cite{wishart2018drugbank} and trial data from \texttt{clinicaltrials.gov}.
For  disease risk, we find historical trial success statistics for different diseases stored in \texttt{clinicaltrials.gov}.

\smallskip
\noindent{\bf Pharmaco-kinetics Knowledge:} We pretrain embeddings using the pharmaco-kinetics (PK) knowledge which is about how body reacts to the intaken drug. 
Because trial success  highly depends on factors such as pharmaco-kinetics properties of a drug and disease risk.  Specifically, we leverage various PK experimental scores generated from the wet labs and stored in various sources on the web.
Utilizing those information, we pretrain on prediction models for Absorption, Distribution, Metabolism, Excretion, Toxicity (ADMET) properties, which are used together in drug discovery to provide insight into how a drug interacts with the body as a whole~\cite{ghosh2016modeling}: 
\begin{itemize}
\item {\bf Absorption:} The Absorption model
describes how drugs can be absorbed into the human body to reach the site of action. A poor absorption drug is usually less desirable. 
\item {\bf Distribution:} The drug Distribution model measures the ability of the molecule to move through bloodstream to various parts of the body. The stronger distribution movement is advisable. 
\item {\bf Metabolism:} The drug metabolism rate determines the duration of a drug's efficacy. 
\item {\bf Excretion: }The drug excretion rate measures how much a drug's toxic components can be removed from the body. 
\item {\bf Toxicity:} The drug toxicity measures damage a drug can cause to the human body.
 \end{itemize}
 
We build predictive models for each of these property scores and its latent embeddings: absorption score $y_A$ and  embedding $\mathbf{h}_A$,  distribution score $y_D$ and embedding $\mathbf{h}_D$,  metabolism score $y_M$ and  embedding $\mathbf{h}_M$, excretion score $y_E$ and embedding  $\mathbf{h}_E$, and  toxicity score $y_T$ and embedding  $\mathbf{h}_T$. 
The input is a molecule while the label is binary, indicating whether the molecule has the desired property. 
\begin{equation}
\label{eqn:admet}
\text{ADMET} \quad
\begin{array}{ll}
\bfh_* = \mathcal{X}_*(\bfh_m), & \bfh_* \in \RB^d, \\
\haty_* = \text{Sigmoid}(\mathrm{FC}(\bfh_{*})), & \haty_* \in [0,1]. \\
\min\ \ \ -y_{*}\log \haty_{*} - (1-y_{*}) \log (1 - \haty_{*}) & y_{*} \in \{0,1\}
\end{array}
\end{equation}

\begin{equation}
\bfh_* = \mathcal{X}_*(\bfh_m) \in \RB^d, \ \   *\in\{A,D,M,E,T\}
\end{equation}

where $\bfh_m \in \RB^d$ is the input drug embedding defined in Eq.~\eqref{eqn:molecule_feature}, $*$ can be A, D, M, E and T. FC is a one-layer fully connected neural network. $\mathcal{X}_{*}$ can be any neural network, we use a multi-layer highway neural network~\cite{srivastava2015highway}.  
We choose highway network because it is able to alleviate the vanishing gradient issue in network training. A single layer of highway network is: 
\begin{equation}
\label{eqn:highway1}
\begin{aligned}
\bfz = T_1(\bfu, \Wf_{T_1}) \odot T_2(\bfu, \Wf_{T_2}) + \bfu \odot (1 - T_2(\bfu, \Wf_{T})),  
\end{aligned}
\end{equation}
where the dimension of $\bfu, \bfz, T_1(\bfu, \Wf_{T_1})$, and $T_2(\bfu, \Wf_{T_2})$ are the same. $\bfu$ and $\bfz$ are input and output for a single layer, respectively. Here $\odot$ is element-wise multiplication, $T_1$ is the affine transform with RELU and $T_2$ is the transform gate with sigmoid.  $T_1$ and $T_2$ are parametrized by $\Wf_{T_1}$ and $\Wf_{T_2}$, respectively.
We have 
\begin{equation}
\label{eqn:highway2}
\bfz = \left\{
\begin{aligned}
 & \bfu, \ &\ \ \ \  \text{if} \ T_2(\bfu,\Wf_{T_2}) = 0, \\
 & T_1(\bfu, \Wf_{T_1}), \ &\ \ \ \  \text{if} \ T_2(\bfu,\Wf_{T_2}) = 1.
\end{aligned}
\right.
\end{equation}
Multiple layers highway network are concatenated. For simplicity, in this paper it is denoted
\begin{equation}
\label{eqn:highway}
\bfz = \text{Highway}(\bfu).  
\end{equation}

Note that the input and output dimensions of highway network are same. If the input embedding is larger than the desired dimension (e.g., concatenation of multiple embedding as input), we first apply a FC layer to reduce the input dimension then apply the highway network in order to maintain the desired dimension. 

\smallskip
\noindent{\bf Disease risk embedding and trial risk prediction:} In addition to drug properties, we also consider the knowledge distilled from historical trials of the target diseases. We consider multiple sources of information about diseases: 1) The disease description and disease ontology, and 2) the historical trial success rate for the disease. As detailed statistics for trial success rate of each disease at different trial phases are widely available~\cite{Hay2014-qw}, we will consider that as the supervision signal to train the {\bf trial risk prediction model}. 
More specifically, given the diseases in the trial, we leverage the historical trial data to predict their success rate, the data is also available at \url{ClinicalTrials.gov}. The predicted trial risk $\haty_{R} $ and embedding $\bfh_{R} \in \RB^d$ are generated via a two-layer highway neural network (Eq.~\ref{eqn:highway}) $\mathcal{R}(\cdot)$:
\begin{equation}
\label{eqn:risk}
\text{Disease Risk}\ \ \ \ \ 
\begin{array}{l}
\bfh_{R} = \mathcal{R}(\bfh_d),   \ \ \ \ \ \ \ \ \ \  \ \ \ \ \ \ \ \ \ \ \ \ \ \ \ \bfh_R \in \RB^d, \\
\haty_{R} = \text{Sigmoid}(\mathrm{FC}(\bfh_{R})),  \ \ \ \ \ \ \ \ \haty_{R} \in [0,1]. \\
\min\ \  -y_{R}\log \haty_{R} - (1-y_{R}) \log (1 -  \haty_{R})    \\
y_R \in \{0,1\}
\end{array}
\end{equation} 
\begin{equation}
\bfh_{R} = \mathcal{R}(\bfh_d) \in \RB^d, 
\end{equation}

where $\bfh_d \in\RB^d$ is the input disease embedding in Eq.~\eqref{eqn:disease_feature}, $\haty_{R}\in[0,1]$ is the predicted trial risk between 0 and 1 (and 0 being the most likely to fail and 1 the most likely to succeed), $y_R \in \{0,1\}$ is the binary label indicating about the success or failure of the trial as a function of disease only. 
and $\mathrm{FC}$ represents the one-layer fully connected layer. 
 Binary cross entropy loss between $y_{R}$ and $\haty_{R}$ is used to guide the training. 

\subsection{Hierarchical Interaction Graph}
\label{sec:gnn}

In this section, we mainly describe (1) the construction of trial interaction graph and (2) how to predict trial outcome using \textit{dynamic attentive graph neural network} on this interaction graph. 

\smallskip
\noindent\textbf{(1) Trial Interaction Graph Construction}
We construct a {\it hierarchical  interaction graph} $\calG$ to connect all input data sources and important factors affecting clinical trial outcomes. Next we describe the interaction graph and its initialization process.

The interaction graph $\calG$ is constructed in a way to reflect the real-world trial development process and it consists of four tiers of nodes that are connected between tiers: 
\begin{itemize}
\item {\bf Input nodes} include drugs, diseases and protocols with node features of input embedding $\bfh_m$, $\bfh_d, \bfh_p \in \RB^d$, colored in green in Figure~\ref{fig:method} (Section~\ref{sec:input}).
\item {\bf External knowledge nodes}  include ADMET embeddings $\bfh_A$, $\bfh_D$, $\bfh_M$, $\bfh_E$, $\bfh_T \in \RB^d$ and disease risk embedding $\bfh_R$. These representation are initialized using pretraining on external knowledge, which are colored blue in Figure~\ref{fig:method} (Section~\ref{sec:knowledge}).
\item {\bf Aggregation nodes} include (1) Interaction node $\bfh_I$ connecting disease $\bfh_d$, drug molecules $\bfh_m$ and protocols $\bfh_p$; (2) pharmaco-kinetics node $\bfh_{PK}$ connecting ADMET embeddings $\bfh_A$, $\bfh_D$, $\bfh_M$, $\bfh_E$, $\bfh_T \in \RB^d$ and (3) augmented interaction node $\bfh_V$ that augment interaction node $\bfh_I$ using disease risk node $\bfh_R$. Aggregation nodes are colored yellow in Figure~\ref{fig:method}. 
\item {\bf Prediction node:} $\bfh_{\text{pred}}$ connect pharmaco-kinetics node $\bfh_{PK}$ and augmented interaction node $\bfh_{V}$ to make the prediction, which is colored in grey in Figure~\ref{fig:method}. 
\end{itemize}
The connection of different nodes are showed in Figure~\ref{fig:method}. Input nodes and external knowledge nodes have been described above and the obtained representation are used as the node embeddings for the interaction graph. Next, we describe aggregation nodes and prediction node.  

\smallskip
\noindent{\bf Aggregation nodes:} The pharmaco-kinetics node is to gather all information of the five ADMET properties (Eq.~\ref{eqn:admet}). We obtain PK embedding by: 
\begin{equation}
\label{eqn:pk}
\text{Pharmaco-Kinetics} \quad \bfh_{PK} = \mathcal{K}(\bfh_A, \bfh_D, \bfh_M, \bfh_E, \bfh_T) \in \RB^d,
\end{equation}
where $\mathcal{K}(\cdot)$ is a one-layer fully-connected layer (input feature is concatenate of $\bfh_A,\bfh_D, \bfh_M, \bfh_E, \bfh_T$, input dimension is $5*d$, output dimension is $d$) followed by $d$-dimensional two-layer highway neural network (Eq.~\ref{eqn:highway})~\cite{srivastava2015highway}.

Then, we model the interaction among the input drug molecule, disease and protocol by an interaction node, where the embedding is obtained by:
\begin{equation}
\label{eqn:interaction}
\text{Interaction}\quad \quad \bfh_I = \mathcal{I}(\bfh_m, \bfh_d, \bfh_p) \in \RB^{d}, 
\end{equation}
where $\bfh_m, \bfh_d, \bfh_p$ are input embeddings defined in Eq.~\eqref{eqn:molecule_feature}, Eq.~\eqref{eqn:disease_feature} and Eq.~\eqref{eqn:protocol_embedding}, respectively. 
The neural architecture of $\mathcal{I}()$ is a one-layer fully connected network (input dimension is $3*d$, output dimension is $d$) followed by $d$-dimensional two-layer highway network (Eq.~\ref{eqn:highway}) \cite{srivastava2015highway}.

Second, we have augmented interaction model to combine (i) trial risk of the target disease $\bfh_R$ (Eq.~\ref{eqn:risk}) and (ii) the interaction among disease, molecule and protocol $\bfh_I$ (Eq.~\ref{eqn:interaction}). 
\begin{equation}
\label{eqn:ai}
\text{Augmented Interaction}\quad \quad \bfh_V = \mathcal{V} ({\bfh_R, \bfh_I}) \in \RB^{d}. 
\end{equation}
The neural architecture of $\mathcal{V}()$ is a one-layer fully connected network (input dimension is $2*d$, output dimension is $d$) followed by $d$-dimensional two-layer highway network (Eq.~\ref{eqn:highway}) \cite{srivastava2015highway}.

\smallskip
\noindent{\bf Prediction node} summarize the pharmaco-kinetics and augmented interaction to obtain the final prediction:
\begin{equation}
\label{eqn:trial}
\text{Trial Prediction}\quad \quad \bfh_{\text{pred}} = \mathcal{P} (\bfh_{PK}, \bfh_{V} )  \in \RB^d 
\end{equation}
Like $\mathcal{I}()$ and $\mathcal{V}()$, the architecture of $\mathcal{P}$ is a one-layer fully connected network (input dimension is $2*d$, output dimension is $d$) followed by $d$-dimensional two-layer highway network (Eq.~\ref{eqn:highway}) \cite{srivastava2015highway}. 

\smallskip
\noindent{\bf (2) Dynamic Attentive Graph Neural Network}
The trial embeddings provide initial representations of different trial components and their interactions via a graph. In order to further enhance predictions, we design a dynamic attentive graph neural network to leverage this interaction graph to model the influential trial components and help improve predictions. 

Mathematically, given the interaction graph $\calG$ as input graph where nodes are trial components and edges are the relations among these trial components. 
We denote $\mathbf{A} \in \{0,1\}^{K\times K}$ as the adjacency matrix of $\calG$. 
The node embeddings $\mathbf{H}^{(0)} \in \RB^{K\times d}$ are initialized to
\begin{equation}
\label{eqn:h0}
\begin{aligned}
\mathbf{H}^{(0)} = [ & \bfh_d, \bfh_m, \bfh_p, \bfh_A, \bfh_D, \bfh_M, \bfh_E, \bfh_T, \bfh_{R}, 
\\ &
\bfh_{PK}, \bfh_{I},  \bfh_{V}, \bfh_{\text{pred}}]^\top \in \RB^{K\times d},
\end{aligned}
\end{equation}
$K = |\calG|$ is number of nodes in graph $\calG$. $K=13$ in this paper. 
We further enhance the node embeddings using graph convolutional network (GCN)~\cite{kipf2016semi}. Eq.~\eqref{eqn:gcn} is the updating rule of GCN for the $l$-th layer, 
\begin{equation}
\label{eqn:gcn}
\begin{aligned}
& \mathbf{H}^{(l)} = \text{RELU}\bigg( \mathbf{B}^{(l)} + (\mathbf{V} \odot \mathbf{A})(\mathbf{H}^{(l-1)}\mathbf{W}^{(l)}) \bigg),  
\\ &
l = 1,\cdots,L, \ \ \ \mathbf{H}^{(l)}\in \RB^{K\times d}, 
\end{aligned}
\end{equation}
where $\mathbf{B} \in \RB^{K\times d}$ is a bias parameter, $\mathbf{W}^{(l)} \in \RB^{d\times d}$ is the weight matrix in $l$-th layer to transform the embedding, $L$ is depth of GCN and $\odot$ is the element-wise multiplication.

Different from conventional GCN~\cite{kipf2016semi}, we introduce a learnable layer-independent \underline{\textit{attentive matrix}} $\mathbf{V}\in \RB^{K\times K}_{+}$. 
$\mathbf{V}_{i,j}$, the ($i,j$)-th entry of $\mathbf{V}$, measures the importance of the edge that connects the $i$-th and $j$-th node in $\calG$. 
We evaluate $\mathbf{V}_{i,j}$ based on the $i$-th and $j$-th nodes' embedding in $\mathbf{H}^{(0)}$, which are denoted $\mathbf{h}^i, \mathbf{h}^j \in \RB^d$, ($\bfh^i\in\RB^d$ is transpose of the $i$-th row of  $\mathbf{H}^{(0)}\in\RB^{K\times d}$ in Eq.~\ref{eqn:h0}) 
\begin{equation}
\label{eqn:attention}
\begin{aligned}
& \mathbf{V}_{i,j} = g_2\big(\text{CONCAT}(\mathbf{h}^i, \mathbf{h}^j)\big), \ \ 
\\ & 
i,j\in\{1,\cdots,K\},\  \mathbf{V}_{i,j}\in \RB_{+}, 
\end{aligned}
\end{equation}
where $g_2(\cdot)$ is a two-layer fully connected neural network with ReLU and Sigmoid activation function in the hidden and output layer respectively. Note that the attentive matrix $\mathbf{V}$ is element-wisely multiplied to the adjacency matrix $\mathbf{A}$ (Eq.~\ref{eqn:gcn}) so that message of edge with higher prediction scores would give a higher weight to propagate. 

\smallskip
\noindent\textbf{Training}
The target is binary label $y \in \{0,1\}$, $y = 1$ indicates the trial succeed while 0 means it fails. 

After GNN message-passing, we obtain an updated representation for each trial components. We then use the last-layer ($L$-th layer) representation on the trial prediction node to generate the trial success prediction as shown in Eq.~\eqref{eqn:predict}, 
\begin{equation}
\label{eqn:predict}
\haty = \text{Sigmoid}(\text{FC}(\bfh^{L}_{\text{pred}})), \ \ \ \haty \in [0,1], 
\end{equation}
where $L$ is the depth of GCN. 
we use one-layer fully-connected network with Sigmoid activation function. Then binary cross entropy loss (in Eq.~\eqref{eqn:bce}) is used to guide the model training, 
\begin{equation}
\label{eqn:bce}
\calL_{\text{classify}} = - y \log \haty - (1-y)\log(1 - \haty). 
\end{equation}
Note that \mname is trained in an end-to-end manner, that is the loss back-propagates to GNN and all the neural networks in Section~\ref{sec:knowledge} and Section~\ref{sec:input} to generate the input representation of node embeddings. 

\begin{algorithm}[h!]
\caption{\mname Framework (with Missing Data Imputation)  } 
\begin{algorithmic}[1]
\STATE \# 1. Pretrain 
\STATE Pretrain basic modules: (i) ADMET models ($\calA, \calD, \calM, \calE, \calT$); (ii) disease risk (DR) model.  
\STATE Construct Interaction Graph $\calG$. 
\STATE \# 2. Train \mname 
\IF{complete data $(\mathbb{M}, \mathbb{D}, \mathbb{C},y)$}
\STATE Fix $\text{IMP}(\cdot)$ (Eq.~\ref{eqn:recovery}), minimize $\calL_{\text{classify}}$ (Eq.~\ref{eqn:bce}), update remaining part of model.
\STATE Minimize $\calL_{\text{recover}}$ (Eq.~\ref{eqn:recovery_loss}), and only update $\text{IMP}(\cdot)$. 
\ELSE
\STATE \#\#\# learn from missing data $(\mathbb{D}, \mathbb{C},y)$. 
\STATE Fix $\text{IMP}(\cdot)$, minimize $\calL_{\text{classify}}$ (Eq.~\ref{eqn:bce}), update remaining part of model. \ \
\ENDIF 
\STATE \# 3. Inference 
\STATE Given new data $(\mathbb{M}, \mathbb{D}, \mathbb{C})$, predict success probability $\haty$. 
\end{algorithmic}
\label{alg:main}
\end{algorithm}

\subsection{Missing Data Imputation}
\label{sec:imputation}

One special challenge associated with trial data is that there can be missing data on molecular information $\mathbb{M}$ due to proprietary information. For complete data, we have $(\mathbb{M},\mathbb{D},\mathbb{P},y)$, whereas for missing data on molecules, we only have $(\mathbb{D},\mathbb{P},y)$. This poses a problem since many nodes representation depend on the molecular information. We observe that there exists high correlation between the drug molecule, disease and protocol features. 
Thus, we design a missing data imputation module based on learning  embeddings that capture inter-modal correlations and intra-modal distribution. In particular, in this study the imputation module $\text{IMP}(\cdot)$ uses disease and protocol embedding ($\bfh_d, \bfh_p$) to recover molecular embedding $\bfh_m$ as given by Eq.~\eqref{eqn:recovery}. 
\begin{equation}
\label{eqn:recovery}
\widehat{\bfh_m} = \text{IMP}(\bfh_d, \bfh_p). 
\end{equation}
Here, we adopt MSE (Mean Square Error) loss as the learning objective to minimize the distance between the ground truth molecule embedding $\bfh_m$ and predicted one $\widehat{\bfh_m}$. 
\begin{equation}
\label{eqn:recovery_loss}
\calL_{\text{recovery}} = \Vert \widehat{\bfh_m} - {\bfh_m}  \Vert_2^2, 
\end{equation}
where $||\cdot||_2$ is $l_2$-norm of a vector. When learning complete data, we update $\text{IMP}(\cdot)$ via minimizing $\calL_{\text{recovery}}$. When learning missing data, we fix $\text{IMP}(\cdot)$, and use $\widehat{\bfh_m}$ to replace $\bfh_m$ and update the remaining part of model. 

For an illustration of the entire framework, we summarize it in Algorithm~\ref{alg:main}.

\section{Benchmark for trial outcome prediction} 
\label{sec:data}

\begin{figure}[t]
\centering
\includegraphics[width = 0.9\textwidth]{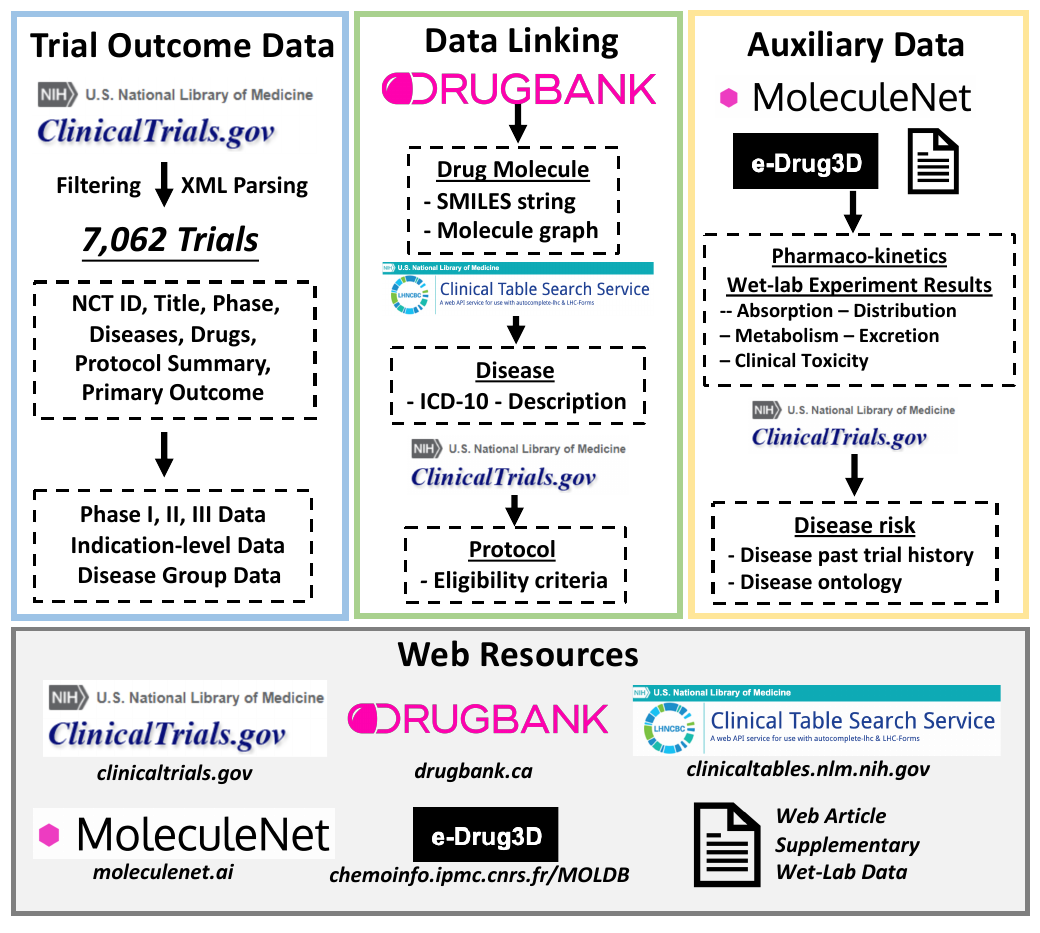}
\caption{Clinical trial outcome benchmark dataset \data. }
\label{fig:benchmark}
\end{figure}

As there is no public trial outcome prediction dataset available, we create a benchmark dataset for Trial Outcome Prediction named \data, which is ready to be released after the double blind review. We first describe the data components and then reports the processing steps to construct this benchmark dataset (Figure~\ref{fig:benchmark}). 
The curated dataset and code repository is available at \footnote{\url{https://github.com/futianfan/clinical-trial-outcome-prediction}}.

\subsection{Benchmark Dataset Overview} 

For each clinical trial, we produce 
1) {\it drug molecule information} including SMILES strings and molecular graphs for the drug candidates used in the trials; 2) {\it disease information} including ICD-10 codes (disease code), disease description and disease hierarchy in terms of CCS codes\footnote{\url{hcup-us.ahrq.gov/toolssoftware/ccs10/ccs10.jsp}} 3) {\it trial protocol information} including eligibility criteria of the trial, study description, outcome measures; 
and 4) {\it trial outcome information} which has a binary indicator of trial success (1) or failure (0), trial phase, start and end date, sponsor, trial size (i.e., number of participants). In addition to the main clinical trial outcome data, we also provide two auxiliary data. One is the pharmaco-kinetics data, which consists of wet lab experiment results for five important PK tasks, along with the drug SMILES strings. Another is disease risk data, which is the past disease trial history success rate and the disease descriptions. 

\smallskip
\noindent{\bf Task.}
There are many tasks that can be studied in term of prediction using \data. In this paper, we focus on trial primary outcome success prediction as a binary classification. Future works can be done for more granular predictions on different types of outcomes such as patient enrollment, expected time and etc. 

\smallskip
\noindent{\bf \data statistics.}
\data consists of 7,062 clinical trials, with 6,483 drugs, 3,820 diseases, 5,582 ICD-10 codes. Out of these trials, 3,448 (48.8\%) succeeded and 3,714 (51.2\%) failed. For pharmaco-kinetics auxiliary dataset, we have 640 drugs for absorption, 1,593 for distribution, 15,020 for metabolism, 15,982 for excretion, 24,576 for toxicity. For disease risk auxiliary dataset, we have 16,356 disease combinations and their success rate in the past.  

\subsection{Data Curation Process} 

We create the \data benchmark for trial outcome prediction from  multiple web sources including drug knowledge base, disease code (ICD-10 code) and historical clinical trials~\cite{anker2016welcome}.

\smallskip
\noindent \textbf{Trial selection.} We apply a series of selection filters to ensure the selected trials have high quality outcome labels. First, we select trials that have associated publications (results or background information) in medical journals as these trials have credible sources. Second, we focus on small molecule drug trials, thus we remove trials with other treatment types such as biologics and behaviorals. Third, we remove observational trials and focus on interventional trials\footnote{\marked{The study type for each record is labelled in \url{ClinicalTrials.gov}}}. Fourth, we select trials that have statistical analysis results for the primary outcome. Each trial in \url{ClinicalTrials.gov} is an XML file and we parse them to obtain the variables. For each trial, we obtain the NCT ID (i.e.,  identifiers to each clinical study), disease names, drugs, brief title and summary, phase, criteria, and statistical analysis results.

\smallskip
\noindent \textbf{Data Processing and Linking.} Next, we describe how do we process and link the parsed trial data to machine learning-ready input and output format: 
\\\noindent$\bullet$  {\it Drug molecule data} are extracted from \url{ClinicalTrials.gov} and linked to its molecule structure (SMILES strings and the molecular graph structures) using DrugBank Database~\cite{wishart2018drugbank} (\url{drugbank.com}). 
\\\noindent$\bullet$ {\it Disease data} are extracted from \url{ClinicalTrials.gov} and linked to ICD-10 codes and disease description using \url{clinicaltables.nlm.nih.gov}  and then to CCS codes via \url{hcup-us.ahrq.gov/toolssoftware/ccs10/ccs10.jsp}. 
\\\noindent$\bullet$  {\it Trial protocol data} are extracted from \url{ClinicalTrials.gov}, in particular, the study description section, outcome section and eligibility criteria section. 
\\\noindent$\bullet$  {\it Trial outcome data} are determined by parsing the study result section on \url{ClinicalTrials.gov}. We use the p-value in the statistical analysis results to induce the primary outcome. The trial is labelled as success if p-value is less than 0.05 and negative if p-value is higher than 0.05. 
\\\noindent$\bullet$ {\it Auxiliary drug pharmaco-kinetics data} include five datasets across the main categories of PK. For absorption, we use the bioavailability dataset provided in Ma et al. paper supplementary~\cite{bioavail}. For distribution, we use the blood-brain-barrier experimental results provided in Adenot et al. study~\cite{BBB}. For metabolism, we use the CYP2C19 experiment from Veith et al.~\cite{CYP} paper, which is hosted in the PubChem biassay portal under AID 1851. For excretion, we use the clearance dataset from the eDrug3D database~\cite{clearance}. For toxicity, we use the ToxCast dataset~\cite{richard2016toxcast}, provided by MoleculeNet (\url{http://moleculenet.ai}). We consider drugs that are not toxic across all toxicology assays as not toxic and otherwise toxic.

\section{Experiments}

\subsection{Experimental Setting}

\noindent\textbf{Implementation Details} 
We use our \data benchmark for model evaluation.
The implementation details are described below. 
The curated dataset and code repository is available at \footnote{\url{https://github.com/futianfan/clinical-trial-outcome-prediction}}. 


\noindent$\bullet$ \textbf{Molecule Embedding} (Section~\ref{sec:input}). Regarding the molecule embedding function $f_m(\cdot)$ in Eq.~\eqref{eqn:molecule_feature}, \mname supports Morgan fingerprint~\cite{cereto2015molecular}, SMILES encoder~\cite{dai2016discriminative}, graph message passing neural network (MPNN)~\cite{jin2018junction,fu2020core,fu2021mimosa,wei2020bitcoin}, we choose MPNN because it usually works better in our experiments. 
The depth of MPNN is 3, with hidden layer dimension 100. The input feature of MPNN is molecular graph with atom and bond features. Following~\cite{jin2018junction,fu2020core,fu2021mimosa}, atom features is a 38 dimensional vector, including its atom type (23 dim, 22 frequent atoms and 1 unknown indicator), degree (6-dim one-hot vector, $\{0,1,2,3,4,5\}$), its formal charge (5-dim one-hot vector, $\{-1,-2,1,2,0\}$) and its chiral configuration (4-dim one-hot vector, $\{0,1,2,3\}$). Bond feature is a 11 dim vector, which is concatenation of its bond type (4-dim one-hot vector, $\{\text{single, double, triple, aromatic}\}$), whether the bond is in a ring (1-dim), and its cis-trans configuration (6-dim one-hot vector, $\{0,1,2,3,4,5\}$). 
\\\noindent$\bullet$ \textbf{Disease Embedding} (Section~\ref{sec:input}). Disease embedding is obtained by GRAM~\cite{choi2017gram,zhao2021pyhealth}, as defined in Eq.~\eqref{eqn:disease_feature}. The embedding size of both $\bfh_d$ (Eq.~\ref{eqn:disease_feature}) and $\bfe$ (Eq.~\ref{eqn:gram}) are both 100. Following \cite{choi2017gram}, the attention model defined in Eq.~\eqref{eqn:gram_attention} is feed-forward network with a single hidden layer (with dimension 100). 
\\\noindent$\bullet$ \textbf{Protocol Embedding} (Section~\ref{sec:input}). $\bfs^I_i$ and $\bfs^E_i$ in Eq.~\eqref{eqn:clinicalbert} are both 768 dimensional vectors.  Following~\cite{gao2020compose}, the kernel size $k_1,k_2,k_3,k_4$ in Eq.~\eqref{eqn:convolutional1} is set to 1,3,5,7, respectively. 
\noindent$\bullet$ \textbf{Interaction Graph} (Section~\ref{sec:gnn}). All the node embedding in Interaction Graph have same dimension to support graph neural network reasoning. The dimension is set to 100. The neural architecture of the connections between different nodes are one-layer fully-connected neural network followed by a two-layer highway network. The input dimension of fully-connected neural network is determined by the number of input node number while the output dimension is 100, equal to the hidden dimension of highway network and node dimension in interaction graph. For example, to obtain pharmaco-kinetics node's embedding in Eq.~\eqref{eqn:pk}, the input dimension of one-layer fully-connected neural network is $500 = 5*100$ ($\bfh_A, \bfh_D, \bfh_M, \bfh_E, \bfh_T$), the output dimension is 100. To obtain Interaction node's embedding in Eq.~\eqref{eqn:interaction}, the input dimension of one-layer fully-connected NN is $300=3*100$ ($\bfh_d, \bfh_m, \bfh_p$). 
\\\noindent$\bullet$ \textbf{Graph Neural Architecture} (Section~\ref{sec:gnn}). 
The hidden size of GCN in Eq.~\eqref{eqn:gcn} is set to 200. The dropout rate is set to 0.6. The feature dimension of GCN is 100, equal to node's embedding size. The graph attention model defined in Eq.~\eqref{eqn:attention} is a two-layer fully-connected neural network with output dimension 1 with sigmoid activation. The input size is $200=2*100$ (concatenation of two nodes' embedding), the hidden size is 50, the output dimension is 1. As shown in Eq.~\eqref{eqn:attention}, the output is a scalar, $\mathbf{V}_{i,j}\in\RB_{+}$. The depth of GNN $L$ is set to 3. 
\\\noindent$\bullet$ \textbf{Learning}. During both pre-training and training procedure, we use the Adam as optimizer~\cite{kingma2014adam}. The learning rate is selected from $\{1e^{-4},5e^{-4},1e^{-3}\}$ and tuned on validation set. When pretraining ADMET models, the learning rate is set to $5e^{-4}$. When pretraining disease risk model, learning rate is set to $1e^{-3}$. When training \mname, the learning rate is set to $5e^{-4}$. The maximal epochs are set to 10. We observe all the models converged within maximal epochs. 
We save the model every epoch and choose the model that performs best on validation set. All hyperparameters are tuned on validation set. 

\smallskip
\noindent\textbf{Evaluation Settings} We consider two realistic evaluation setups. The first is {\bf phase-level evaluation} where we predict the outcome of a single phase study. Since each phase has different goals (e.g. Phase I is for safety whereas Phase II and III is for efficacy), we conduct evaluation for Phase I, II and III individually. We create the test datasets using the FDA guideline\footnote{https://www.fda.gov/patients/drug-development-process/step-3-clinical-research} on the success-failure ratio for each phase, specifically 70\% success rate for Phase I, 33\% for Phase II and 30\% for Phase III. Second, we also consider {\bf indication-level evaluation} where we test if the drug can pass all three phases for the final market approval. To imitate it, we assemble all phase studies related to the drug and disease of the study and then use the latest phase protocol as the input to our model. Drugs that have Phase III success are labelled positive and other drugs that fail in any of the three phases are labelled negative. Data statistics are shown in Table~\ref{table:statistics}.

\begin{table}[h!]
\small 
\centering
\caption{Data statistics. During training, we randomly select 15\% training samples for model validation. \marked{The earlier trials are used for learning while the later trials are used for inference.} }
\label{table:statistics}
\begin{tabular}{l|ccccc}
\toprule
\multirow{2}{*}{Settings}  & \multicolumn{2}{c}{Train} & \multicolumn{2}{c}{Test} & \multirow{2}{*}{\marked{Split Date}} \\
& Success & Failure & Success & Failure &  \\ \midrule 
Phase I  & 702 & 386 & 199 & 113  &  Aug 13, 2014 \\
Phase II  & 956 & 1655 & 302 & 487 & March 20, 2014 \\
Phase III  & 1,820 & 2,493 & 457 & 684 & April 7, 2014 \\
Indication & 1,864 & 2,922 & 473 & 674 & May 21, 2014 \\
\bottomrule
\end{tabular}
\end{table}

\smallskip
\noindent\textbf{Evaluation Metrics} 
We use the following metrics to measure the performance of all methods. 
\\\noindent$\bullet$  \textbf{PR-AUC} (Precision-Recall Area Under Curve). 
\\\noindent$\bullet$  \textbf{F1}. The F1 score is the harmonic mean of the precision and recall.
\\\noindent$\bullet$  \textbf{ROC-AUC} (Area Under the Receiver Operating Characteristic Curve). 
\\\noindent$\bullet$  \marked{\textbf{p-value}. We report the results of hypothesis testing in terms of p-value to showcase the statistical significance of our method over the best baseline results. If p-value is smaller than 0.05, we claim our method significantly outperforms the best baseline method. } 

For PR-AUC, F1, and ROC-AUC, higher value represent better performance. 
\marked{We split the dataset based on registration date. The earlier trials are used for learning while the later trials are used for inference. For example, for Phase I dataset, we learn the model using the trials before Aug 13th, 2014 and make inference using the trials after that date, as shown in Table~\ref{table:statistics}.  
We use Bootstrap~\cite{chernick2011bootstrap} to estimate the mean and standard deviation of accuracy on test set. 
}

\smallskip
\noindent\textbf{Baselines}. We compare \mname with several baselines, including both conventional machine learning models and deep learning methods. 
\begin{itemize}[leftmargin=*]
 \item \textbf{LR} (Logistic Regression). It was used in \cite{Lo2019Machine} on trial prediction with disease features only. For fair comparison, we adapt it such that the input features include (i) 1024 dimensional Morgan fingerprint feature~\cite{cereto2015molecular}, (ii) GRAM embedding (Eq.~\ref{eqn:gram}, GRAM is pretrained using disease risk module~Eq.~\ref{eqn:risk}) and (iii) BERT embedding of eligibility criteria for protocol. Then these three features are concatenated as the input of LR model. 
 \item \textbf{RF} (Random Forest). Similarly to LR, it was used in \cite{Lo2019Machine} on trial prediction and we adapt it to use the same feature set. 
 \item \textbf{XGBoost}. An implementation of gradient boosted decision trees designed for speed and performance. It was used in context of individual patient trial outcome prediction in~\cite{raj20evaluation}. We adapt it to use the same feature set for general trial outcome prediction. 
 \item \textbf{AdaBoost} (Adaptive Boosting). It was used in~\cite{fan2020application} for individual Alzheimer’s patient's trial result prediction. We adapt it to use the same feature set. 
 \item \textbf{kNN+RF}. \cite{Lo2019Machine} leverages statistical imputation techniques to handle missing data, and finds using (1) kNN (k-Nearest Neighbor) as imputation technique and (2) Random Forest as classifier would achieve best performance. We adapt this method to use the same feature set. 
\item \textbf{FFNN} (Feed-Forward Neural Network)~\cite{Tranchevent750364}. It uses the same feature with LR. The feature vectors are fed into a three-layer feedforward neural network. 
\item \textbf{DeepEnroll}~\cite{zhang2020deepenroll}. DeepEnroll was originally designed for patient trial matching, it uses (1) pre-trained BERT model~\cite{devlin2018bert} to encode eligibility criteria into sentence embedding; (2) a hierarchical embedding model to disease information and (3) alignment model to capture the protocol-disease interaction information. To adapt it to our scenario, molecule embedding ($\bfh_m$) is concatenated to the output of alignment model to make prediction. 
\item \textbf{COMPOSE} (cross-modal pseudo-siamese network)~\cite{gao2020compose}. COMPOSE was also originally designed for patient trial matching, it uses convolutional highway network and memory network to encode eligibility criteria and diseases respectively and an alignment model to model the interaction. COMPOSE incorporate the molecule information in the same way with DeepEnroll, as mentioned above. 
\end{itemize}

\subsection*{Exp 1. Phase Level Trial Outcome Prediction}

\begin{table}[h!]
\small 
\centering
\caption{Empirical results of various approaches for \textbf{Phase Level} outcome prediction. The mean and standard deviation are reported. }
\label{table:phase_result}
\begin{tabular}{lccc}
\toprule
\multicolumn{4}{c}{\bf Phase I Trials} \\ 
 Method & PR-AUC & F1 & ROC-AUC  \\
\midrule
\marked{LR} & 0.575{\stdfontsize $\pm$0.011} & 0.640{\stdfontsize $\pm$0.013} & 0.630{\stdfontsize $\pm$0.017} \\
\marked{RF} & 0.640{\stdfontsize $\pm$0.012} & 0.656{\stdfontsize $\pm$0.017} &  0.674{\stdfontsize $\pm$0.013} \\ 
\marked{XGBoost} & 0.653{\stdfontsize $\pm$0.015} & 0.671{\stdfontsize $\pm$0.016} & 0.698{\stdfontsize $\pm$0.012} \\ 
\marked{AdaBoost} & 0.589{\stdfontsize$\pm$0.010} & 0.612{\stdfontsize$\pm$0.015} & 0.623{\stdfontsize$\pm$0.012}  \\ 
\marked{kNN+RF}~\cite{Lo2019Machine} & 0.616{\stdfontsize$\pm$0.016} & 0.625{\stdfontsize$\pm$0.021} & 0.624{\stdfontsize$\pm$0.015}  \\ 
\marked{FFNN}~\cite{Tranchevent750364} & 0.643{\stdfontsize$\pm$0.020} & 0.745{\stdfontsize$\pm$0.024} & 0.747{\stdfontsize$\pm$0.026}  \\ 
\marked{DeepEnroll}~\cite{zhang2020deepenroll} & 0.654{\stdfontsize$\pm$0.020} & 0.754{\stdfontsize$\pm$0.019}  & 0.750{\stdfontsize$\pm$0.021}  \\ 
\marked{COMPOSE}~\cite{gao2020compose}  & 0.681{\stdfontsize$\pm$0.017}  & 0.768{\stdfontsize$\pm$0.019}  & 0.766{\stdfontsize$\pm$0.016}  \\\midrule
\marked{\mname - Pretrain} & 0.701{\stdfontsize$\pm$0.022} & 0.792{\stdfontsize$\pm$0.018}  & 0.784{\stdfontsize$\pm$0.020}    \\
\marked{\mname - GNN} & 0.753{\stdfontsize$\pm$0.020} & 0.819{\stdfontsize$\pm$0.014} & 0.813{\stdfontsize$\pm$0.016} \\
\marked{\mname} &  \bf 0.764 {\stdfontsize $\pm$0.018}  & \bf 0.837{\stdfontsize $\pm$0.012}  & \bf 0.817{\stdfontsize $\pm$0.015} \\ 
\midrule  \marked{p-value} & 0.00003 & 0.0001 & 0.0004  \\
\midrule
\multicolumn{4}{c}{\bf Phase II Trials} \\ 
 Method & PR-AUC & F1 & ROC-AUC  \\
\midrule
\marked{LR} & 0.489{\stdfontsize$\pm$0.012} & 0.528{\stdfontsize$\pm$0.018} & 0.600{\stdfontsize$\pm$0.016} \\ 
\marked{RF} & 0.578{\stdfontsize$\pm$0.020} & 0.611{\stdfontsize$\pm$0.019} & 0.683{\stdfontsize$\pm$0.022}  \\ 
\marked{XGBoost} & 0.571{\stdfontsize$\pm$0.020} & 0.608{\stdfontsize$\pm$0.012} & 0.694{\stdfontsize$\pm$0.013} \\ 
\marked{AdaBoost} & 0.457{\stdfontsize$\pm$0.014} & 0.515{\stdfontsize$\pm$0.017} & 0.556{\stdfontsize$\pm$0.013} \\ 
\marked{kNN+RF}~\cite{Lo2019Machine} & 0.544{\stdfontsize$\pm$0.018} & 0.560{\stdfontsize$\pm$0.019} & 0.667{\stdfontsize$\pm$0.014} \\ 
\marked{FFNN}~\cite{Tranchevent750364}  & 0.555{\stdfontsize$\pm$0.020} & 0.569{\stdfontsize$\pm$0.028} & 0.661{\stdfontsize$\pm$0.021} \\
\marked{DeepEnroll}~\cite{zhang2020deepenroll} & 0.560{\stdfontsize$\pm$0.018} & 0.598{\stdfontsize$\pm$0.020} & 0.742{\stdfontsize$\pm$0.017} \\ 
\marked{COMPOSE}~\cite{gao2020compose}  & 0.570{\stdfontsize$\pm$0.017} & 0.612{\stdfontsize$\pm$0.015} & 0.743{\stdfontsize$\pm$0.016}  \\ 
\midrule
\marked{\mname - Pretrain} & 0.583{\stdfontsize$\pm$0.017} & 0.621{\stdfontsize$\pm$0.017} & 0.757{\stdfontsize$\pm$0.018} \\ 
\marked{\mname - GNN} & 0.595{\stdfontsize$\pm$0.018} & 0.626{\stdfontsize$\pm$0.017} & 0.751{\stdfontsize$\pm$0.018} \\ 
\marked{\mname} & \bf 0.602{\stdfontsize$\pm$0.016}  & \bf 0.638{\stdfontsize$\pm$0.015} & \bf 0.761{\stdfontsize$\pm$0.018} \\ 
\midrule  \marked{p-value} & 0.008 & 0.012 &  0.048 \\
\midrule
\multicolumn{4}{c}{\bf Phase III Trials} \\ 
 Method & PR-AUC & F1 & ROC-AUC \\
\midrule
\marked{LR} & 0.533{\stdfontsize$\pm$0.005} & 0.590{\stdfontsize$\pm$0.009} & 0.642{\stdfontsize$\pm$0.007} \\
\marked{RF} & 0.554{\stdfontsize$\pm$0.008} & 0.626{\stdfontsize$\pm$0.010} & 0.680{\stdfontsize$\pm$0.011}  \\
\marked{XGBoost} & 0.588{\stdfontsize$\pm$0.013} & 0.621{\stdfontsize$\pm$0.014} & 0.719{\stdfontsize$\pm$0.015}  \\ 
\marked{AdaBoost} & 0.560{\stdfontsize$\pm$0.012} & 0.590{\stdfontsize$\pm$0.015} & 0.678{\stdfontsize$\pm$0.014}  \\ 
\marked{kNN+RF}~\cite{Lo2019Machine} & 0.560{\stdfontsize$\pm$0.012} & 0.601{\stdfontsize$\pm$0.013} & 0.668{\stdfontsize$\pm$0.016} \\ 

\marked{FFNN}~\cite{Tranchevent750364} & 0.576{\stdfontsize$\pm$0.018} & 0.628{\stdfontsize$\pm$0.020} & 0.684{\stdfontsize$\pm$0.018} \\
\marked{DeepEnroll}~\cite{zhang2020deepenroll} & 0.581{\stdfontsize$\pm$0.016} & 0.646{\stdfontsize$\pm$0.020} & 0.699{\stdfontsize$\pm$0.016} \\ 
\marked{COMPOSE}~\cite{gao2020compose}  & 0.589{\stdfontsize$\pm$0.012} & 0.653{\stdfontsize$\pm$0.016} & 0.715{\stdfontsize$\pm$0.016} \\   
\midrule
\marked{\mname - Pretrain}  & 0.599{\stdfontsize$\pm$0.018} & 0.658{\stdfontsize$\pm$0.014} & 0.721{\stdfontsize$\pm$0.017}  \\ 
\marked{\mname - GNN} & \bf 0.622{\stdfontsize$\pm$0.014} & \bf 0.693{\stdfontsize$\pm$0.018} & \bf 0.759{\stdfontsize$\pm$0.018}  \\
\marked{\mname} &  0.618{\stdfontsize$\pm$0.014}  &  0.683{\stdfontsize$\pm$0.016}  &  0.726{\stdfontsize$\pm$0.015}  \\ 
\midrule  \marked{p-value} & 0.002 & 0.009 & 0.14  \\
\bottomrule
\end{tabular}
\end{table}

First, we conduct phase level outcome prediction. For each phase, we train a separate model to make the prediction. 
We compare \mname with several baseline approaches, which cover conventional machine learning models and deep learning based models. We control the ratio of learning/inference data number to about 4:1, during learning, we leave 15\% training data as validation set. The means and standard deviations of the results are reported. 
We present the prediction performance in Table~\ref{table:phase_result}. 
We have the following observations: 

(1) Deep learning based approaches including FFNN, DeepEnroll, COMPOSE and \mname outperforms conventional machine learning approaches (LR, RF, XGBoost, AdaBoost, kNN+RF) significantly in outcome prediction for all the three phases, thus validating the benefit of deep learning methods for clinical trial outcome prediction. 

(2) Among all the deep learning methods, \mname performs best with 0.772 PR-AUC for phase I, 0.607 for phase II and 0.623 for phase III.
Compared with the strongest baseline (COMPOSE), which is also deep learning approach that uses all the features, \mname achieved 12.4\%, 3.5\%, 3.0\% relatively improvement in terms of PR-AUC and 8.2\%, 2.7\%, 6.5\% relative improvement in terms of F1 score. 
The reason is that \mname incorporates insightful multimodal data embedding and finer-grained interaction between multimodal data and trial components (i.e., node in interaction graph). 

(3) The full \mname performs better than the variant without pre-trained model (\mname - Pretrain) and  the one without using GNN model (\mname - GNN) in both phase I and II scenario. This observation confirmed the importance of all modules in \mname.

(4) When comparing the prediction performance across phase I, II and III, we find that phase I achieves highest accuracy for almost all the methods while phase II is most challenging with lowest accuracy. 
This result is consistent with historical trials statistics~\cite{thomas2016clinical} and reported accuracy of machine learning models on these tasks~\cite{Lo2019Machine}. 


\subsection*{Exp 2. Indication Level Outcome Prediction}
The indication level outcome prediction focuses on predicting whether a trial will pass all three phases. We build a separate model using a combined dataset where the successful trials are the ones passed phase III (plus the ones reached phase IV) while the failed trials are the ones that failed in any phase from I to III. 
The results are presented in Table~\ref{table:indication}. 
Similar trends are observed. In particular, \mname performs the best with 0.703 PR-AUC, 0.765 F1 and 0.793 ROC-AUC which achieves 8.2\%, 5.7\%, 1.2\% relative improvements on PR-AUC, F1 and ROC-AUC over the strongest baseline (COMPOSE).

\begin{table}[h!]
\small 
\centering
\caption{Performance on indication level  prediction. The mean and standard deviation are reported.}
\label{table:indication}
\begin{tabular}{lccc}
\toprule
 Method & PR-AUC & F1 & ROC-AUC  \\
\midrule
\marked{LR} & 0.579{\stdfontsize$\pm$0.007} & 0.613{\stdfontsize$\pm$0.010} & 0.645{\stdfontsize$\pm$0.009}  \\ 
\marked{RF} & 0.594{\stdfontsize$\pm$0.012} & 0.627{\stdfontsize$\pm$0.012} & 0.621{\stdfontsize$\pm$0.011}  \\ 
\marked{XGBoost} & 0.603{\stdfontsize$\pm$0.010} & 0.614{\stdfontsize$\pm$0.012} & 0.645{\stdfontsize$\pm$0.011} \\
\marked{AdaBoost} & 0.565{\stdfontsize$\pm$0.007} & 0.597{\stdfontsize$\pm$0.011} & 0.632{\stdfontsize$\pm$0.011} \\
\marked{kNN+RF}~\cite{Lo2019Machine} & 0.594{\stdfontsize$\pm$0.010} & 0.643{\stdfontsize$\pm$0.014} & 0.700{\stdfontsize$\pm$0.011} \\
\marked{FFNN}~\cite{Tranchevent750364} & 0.602{\stdfontsize$\pm$0.017} & 0.673{\stdfontsize$\pm$0.019} & 0.708{\stdfontsize$\pm$0.014}   \\ 
\marked{DeepEnroll}~\cite{zhang2020deepenroll} & 0.616{\stdfontsize$\pm$0.016} & 0.675{\stdfontsize$\pm$0.015} & 0.712{\stdfontsize$\pm$0.013}    \\ 
\marked{COMPOSE~\cite{gao2020compose}}  & 0.629{\stdfontsize$\pm$0.017} & 0.720{\stdfontsize$\pm$0.015} & 0.732{\stdfontsize$\pm$0.014} \\\midrule
\marked{\mname - Pretrain}  & 0.642{\stdfontsize$\pm$0.015} & 0.731{\stdfontsize$\pm$0.016} & 0.749{\stdfontsize$\pm$0.018}   \\ 
\marked{\mname - GNN} &  0.653{\stdfontsize$\pm$0.014} & 0.756{\stdfontsize$\pm$0.013} & 0.774{\stdfontsize$\pm$0.014}   \\ 
\marked{\mname} &  \bf 0.670{\stdfontsize$\pm$0.014}  &  \bf 0.762{\stdfontsize$\pm$0.012}  & \bf  0.783{\stdfontsize$\pm$0.017}  \\ 
\midrule  \marked{p-value} & 0.002 & 0.001 &  0.0004 \\  
\bottomrule
\end{tabular}
\end{table}

\section{Conclusion}
In this paper, we propose the a general clinical trial outcome prediction task. We design a \fullname (\mname) to leverage multi-sourced data and incorporate multiple factors in a hierarchical interaction graph. Also, it can handle missing data via imputation module. Empirical studies indicate that \mname outperforms baseline methods in several metrics, obtaining state-of-the-art predictive measures on indication-level and phase level outcome prediction.

\bibliographystyle{plain}
\bibliography{ref}




\end{document}